\documentclass[fleqn,usenatbib]{mnras}

\usepackage{newtxtext,newtxmath}

\usepackage[T1]{fontenc}
\usepackage{ae,aecompl}
\usepackage{multirow}

\usepackage{graphicx}	
\usepackage{amsmath}	
\usepackage{hyperref}
\usepackage{footmisc} 

\usepackage{booktabs}
\usepackage{xspace}
\usepackage{verbatim}

\newcommand{\apx}[1]{^{\rm #1}}
\newcommand{\pdx}[1]{_{\rm #1}}

\def\nh{$N_{\rm H}$\xspace}
\def\nhmax{$N_{\rm H,~tot}$\xspace}
\def\col{cm$^{-2}$\xspace}

\def\flux{erg~s$^{-1}$~cm$^{-2}$\xspace}
\def\lum{erg~s$^{-1}$\xspace}

\def\deg{^\circ}

\def\hr{HR$_1$--HR$_2$\space}

\def\xmm{{\em XMM-Newton}\xspace}
\def\dr{4XMM-DR10\xspace}

\def\swift{{\em Swift}\xspace}
\def\cha{{\em Chandra}\xspace}

\usepackage{color}

\title[Candidate INSs in \dr]{Candidate isolated neutron stars in the \dr catalog of X-ray sources}

\author[Rigoselli et al.]{Michela Rigoselli$^{1}\thanks{E-mail: michela.rigoselli@inaf.it}$, Sandro Mereghetti$^{1}$, Caterina Tresoldi$^{1,2}$
\\
$^{1}$INAF, Istituto di Astrofisica Spaziale e Fisica Cosmica Milano, via A.\ Corti 12, I-20133 Milano, Italy\\
$^{2}$Dipartimento di Fisica, Università degli Studi di Milano, Via Celoria 16, I-20133 Milano, Italy
}

\date{Accepted 2021 October 12. Received 2021 October 12; in original form 2021 July 31}

\pubyear{2021}

\begin{document}
\label{firstpage}
\pagerange{\pageref{firstpage}--\pageref{lastpage}}
\maketitle

\begin{abstract}
Most isolated neutron stars have been discovered thanks to  the detection of their pulsed non-thermal emission, at wavelengths spanning from radio to gamma-rays.  However, if the beamed non-thermal radiation does not intercept our line of sight or it is too faint or absent, isolated neutron stars can also be detected through their thermal emission, which peaks in the soft X-ray band and is emitted nearly isotropically. 
In the past thirty years, several thermally-emitting isolated neutron stars have been discovered thanks to X-ray all-sky surveys, observations targeted at the center of supernova remnants,  or as serendipitous X-ray sources. Distinctive properties of these relatively rare X-ray sources are very soft spectra and high ratios of X-ray to optical flux. 
The recently released \dr catalog contains more than half a million X-ray sources detected with  the \xmm telescope in the 0.2--10 keV range in observations carried out from 2000 to 2019.
Based on a study of the spectral properties of these sources and on cross-correlations with catalogs of possible counterparts, we have carried out a search of isolated neutron stars, finding four potential candidates. The spectral and long-term variability analysis of these candidates, using also \cha and \swift-XRT data, allowed us to point out the most interesting sources deserving further multiwavelength investigations.
\end{abstract}


\section{Introduction}

Neutron stars are the remnants of massive stars whose cores collapse during the supernova explosions. They are usually detected as radio ($\sim$2500 known,  \citealt{man05}) and/or gamma-ray ($\sim$300, \citealt{4fermi}) pulsars, thanks to their  beamed non-thermal emission.
In addition, there is a  group of isolated neutron stars (INSs) from which pulsed non-thermal emission is not detected. 

This could be caused by an unfavorable orientation of the rotation and magnetic axes, or by the intrinsic lack, or faintness, of non-thermal magnetospheric emission.
On the other hand, these objects can be discovered through their thermal X-ray emission, that arises from the cooling neutron star surface and, being emitted nearly isotropically, can be seen independently of the star orientation. This thermal emission is characterized by a soft X-ray spectrum that peaks between 0.2 and 2 keV \citep[and references therein]{pot20} and gives only a small contribution in the optical band, leading to high values of the X-ray to optical flux ratio, $F\pdx{X}/F\pdx{O} \gtrsim 10^{3}$.
Pulsations with broad, nearly sinusoidal profiles, and usually small pulsed fractions  (e.g., \citealt{2007ApJ...657L.101T}), can be detected. These are likely caused by a non-uniform   surface temperature and/or beaming effects due to the presence of a magnetized atmosphere.

The prototype of these thermally emitting INSs was discovered with the ROSAT satellite \citep{1996Natur.379..233W}, and, subsequently, other six similar sources were found in the ROSAT All-Sky Survey (RASS, \citealt{1999A&A...349..389V}). These sources, generally referred to as XDINS (X-ray dim isolated neutron stars), have distances in the range 100--500 pc, X-ray spectra with blackbody temperatures of 45--110 eV and spin periods of a few seconds \citep{2007Ap&SS.308..181H,2009ASSL..357..141T}.

Thanks to their high sensitivity, X-ray satellites such as \xmm and \cha should be able to detect dimmer (and thus farther) thermally emitting INSs. However, they provide a smaller sky coverage than all-sky surveys. Furthermore, absorption by the interstellar medium strongly suppresses soft X-rays, thus reducing the possibility of detecting farther and more absorbed objects.
In fact, further searches using these satellites led only to a small increase of this sample. For example, \citet{2009A&A...504..185Pires} analyzed the 2XMMp catalog of serendipitous sources discovered with \xmm and found a few possible INS candidates, among which the most promising is 2XMM J104608.7$-$594306 \citep{2009A&A...498..233Pires2,2015A&A...583A.117P}.

Based on a population synthesis model accounting for the distribution of neutron star birthplaces and of the local ($<$3 kpc) interstellar absorption, \citet{2008A&A...482..617P}, suggested that the most promising regions to look for INSs are in the direction of rich OB associations.  
The eROSITA/SRG instrument \citep{2021A&A...647A...1P} has recently completed the first of the planned four years of its  all sky survey. This has provided the first coverage of the whole sky carried out in the soft X-rays after the RASS,  although in a slightly different energy range: 0.1--10 keV  wrt 0.1--2.4 keV.
Predictions for the expected number of INS in the complete eROSITA survey have been reported by \citet{2017AN....338..213P}, who estimate about 90 discoveries, down to 0.2--2 keV fluxes of $\sim$10$^{-14}$ \flux in the whole sky. However, their secure identification in the large sample of eROSITA sources will require extensive multiwawelength follow-up observations.

Despite these difficulties, the recent release of the \dr catalog, containing more than half a million X-ray sources seen with \xmm,  motivated us to perform a new search for INS candidates.
This paper is organized as follows. In Section~\ref{sec:hr} we describe how we characterized the X-ray spectrum of the INS in a suitable way to browse the \xmm source catalog.
We then describe the filtering process to exclude extended and variable sources, and those with a bright optical and/or IR counterpart (Section~\ref{sec:catalog}), and finally we analyze the new possible INSs candidates (Section~\ref{sec:ss}). The results are discussed in Section~\ref{sec:disc}.

\section{X-ray hardness ratios of \dr sources}\label{sec:hr}

The \dr source catalog was released on 2020 December 10 \citep{2020A&A...641A.136W}. It contains $849,991$ source detections drawn from $11,647$ \xmm EPIC observations obtained in the first 20 years of satellite operations. The observations cover $\sim$1192 deg$^2$ of the sky, and contain $575,158$ unique sources.

For each detection, the catalog lists many parameters derived by the pipelines of the Science Analysis System (SAS)\footnote{http://www.cosmos.esa.int/web/xmm-newton/sas.}. These include the sky coordinates, the extension of the source, some flags indicating the quality of the detection and the possible variability, and the net (background-subtracted) count rates in five energy bands: (1) 0.2--0.5 keV,  (2) 0.5--1 keV,  (3) 1--2 keV,  (4) 2--4.5 keV, and (5) 4.5--12 keV. 

A convenient way to characterize the X-ray spectrum of a source when the available number of counts is too small for a spectral fitting, is to look at the X-ray hardness ratios (HRs). They are defined as
\begin{equation}
\mathrm{HR}_{i} = \frac{\mathrm{CR}_{i+1}-\mathrm{CR}_{i}}{\mathrm{CR}_{i+1}+\mathrm{CR}_{i}} \qquad i=1,\dots4
\label{eq:hr}
\end{equation}
where CR$_{i}$ and CR$_{i+1}$ are the count rates in two adjacent energy bands. Soft sources are best described by HR$_1$ and HR$_2$.

\begin{figure}
  \centering
  \includegraphics[width=0.5\textwidth]{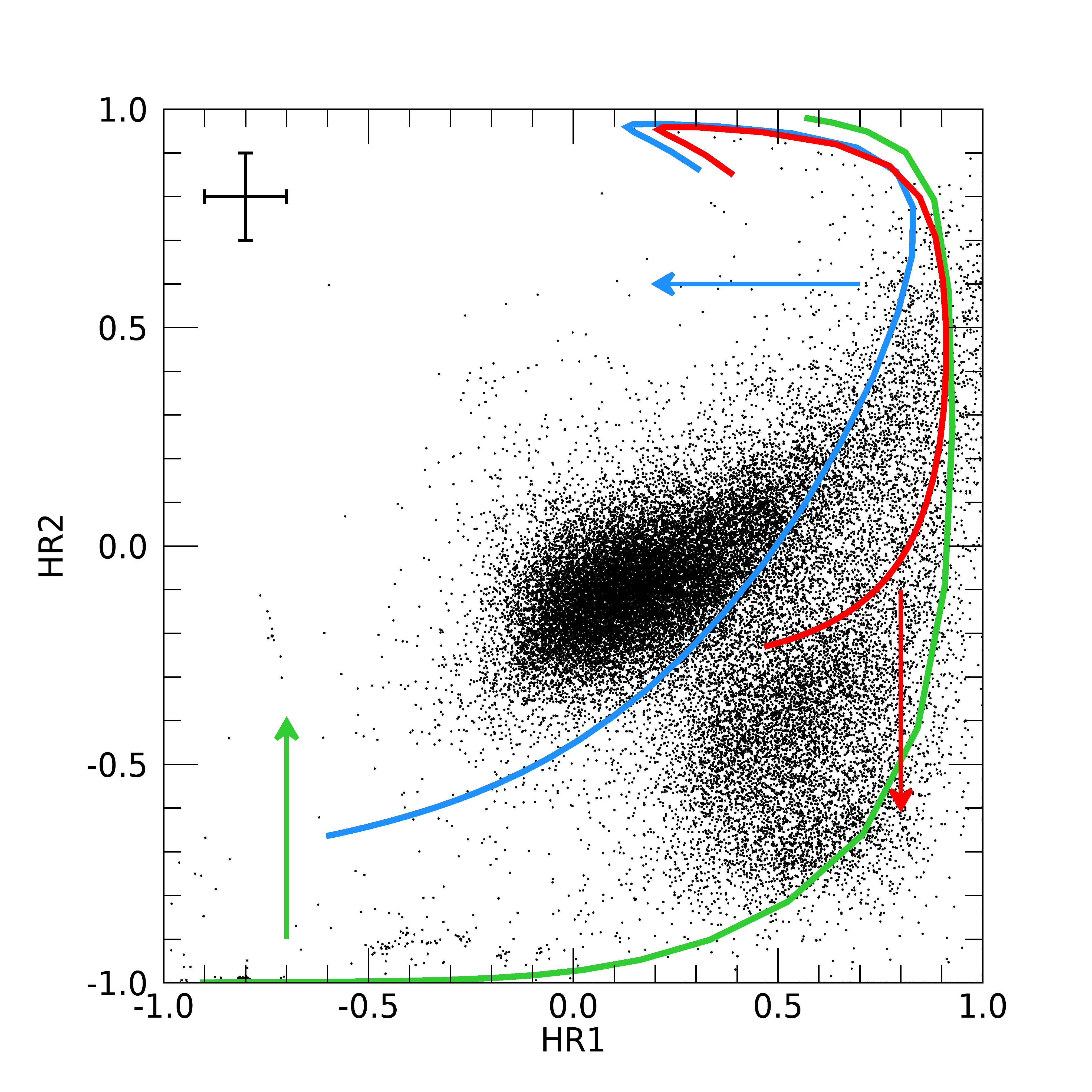}
  \caption{\hr plot of the the point-like \dr sources with errors on the hardness ratios  smaller than 0.1 (the error bars in the top left corner illustrate this uncertainty). The sources cluster into two main groups: The AGNs, having a non-thermal spectrum, are placed to the left of the blue line (\textsc{powerlaw}, $\Gamma\lesssim4$); the stars, having a thermal spectrum, are placed in the region below the red line indicated by the arrow (\textsc{apec}, $kT\lesssim1.05$ keV, 0.5 solar abundance). The green line represents a blackbody (\textsc{bbodyrad}) 
  of temperature $kT=0.05$ keV; typical INSs and SSSs, with temperatures in the range 0.05--0.25 keV, can be found in the region above this line indicated by the green arrow. The lines are obtained varying \nh from $0$ to $10^{23}$ \col using the \textsc{tbabs} model.}
  \label{fig:hr}
\end{figure}

Figure~\ref{fig:hr} shows the HR$_1$ and HR$_2$ values of \dr point-like  sources (only sources with HR errors $\leq$0.1 are plotted). The majority of the sources are clearly grouped in two distinct regions. It has been shown in previous works (see e.g. \citealt{2012ApJ...756...27L}) that the active galactic nuclei (AGNs, comprising  BL Lac, Blazars, Quasars...) are located in the center of the \hr plane, while the lower right  corner is mostly populated by stars. The blue line corresponds to the expected HR values (see below) for power-law spectra of photon index $\Gamma = 4$ and different values of absorption. AGNs, which typically have $\Gamma < 4$, lie to the left of the blue line. The spectra of stars can be described by hot plasma thermal models. The red line corresponds to emission from a  collisionally ionized gas with temperature $kT = 1.05$ keV and different absorption values. Most stars have a lower temperature and lie below this line. 

Compact objects form a third group of sources, comprising  INSs, supernova remnants (SNRs) (that appear point-like when located in other galaxies), and  binary systems hosting a white dwarf, a neutron star or a stellar mass black hole. Their spectral and timing properties can be used to distinguish several classes \citep{2006ARA&A..44..323F}: The super-soft sources (SSSs), cataclysmic variables (CVs), novae, high-mass and low-mass X-ray binaries (HMXBs and LMXBs), ultraluminous X-ray sources (ULXs). The group of compact objects does not have a specific location on the \hr plane. However, in the lower left corner of the plane (HR$_1\lesssim0$ and HR$_2\lesssim-0.5$), there are only cold INSs and SSSs (the green line represents a blackbody with $kT=0.05$ keV, and hotter blackbodies are above it).

Knowing the spectral response of the X-ray detector, it is possible to compute the expected HR values for any specific source emission model and value of the interstellar absorption. 
For single-component models, one can fix the  parameter (e.g. the  photon index or the temperature) and vary the  absorption to obtain tracks on the \hr plane, such as those shown by the lines in  Figure~\ref{fig:hr}. 

In order to derive HR values appropriate to select thermally emitting INSs, we considered a blackbody model with temperature $kT\in$ $[0.05,0.25]$ keV and absorption column density \nh~$\in$ $[10^{20},10^{22}]$ \col.
We used the models implemented in XSPEC, and the interstellar absorption with the Tuebingen-Boulder ISM absorption cross section, with abundances from \citet{2000ApJ...542..914W}.
Folding  the model with the response functions of the EPIC-pn camera for different operating modes, optical filters and off-axis angles, we computed the expected count rates in the five standard energy bands defined above. We found significant differences in the resulting HR values only as a function of the optical filters. In fact, the HRs computed for different observing modes and off-axis angles differ by less than 0.05. Therefore, in the following we adopted, for each of the three filters (thin, medium and thick) the corresponding HR values computed with the on-axis Full Frame response matrices. 

\begin{figure*}
  \centering
  \includegraphics[width=0.49\textwidth]{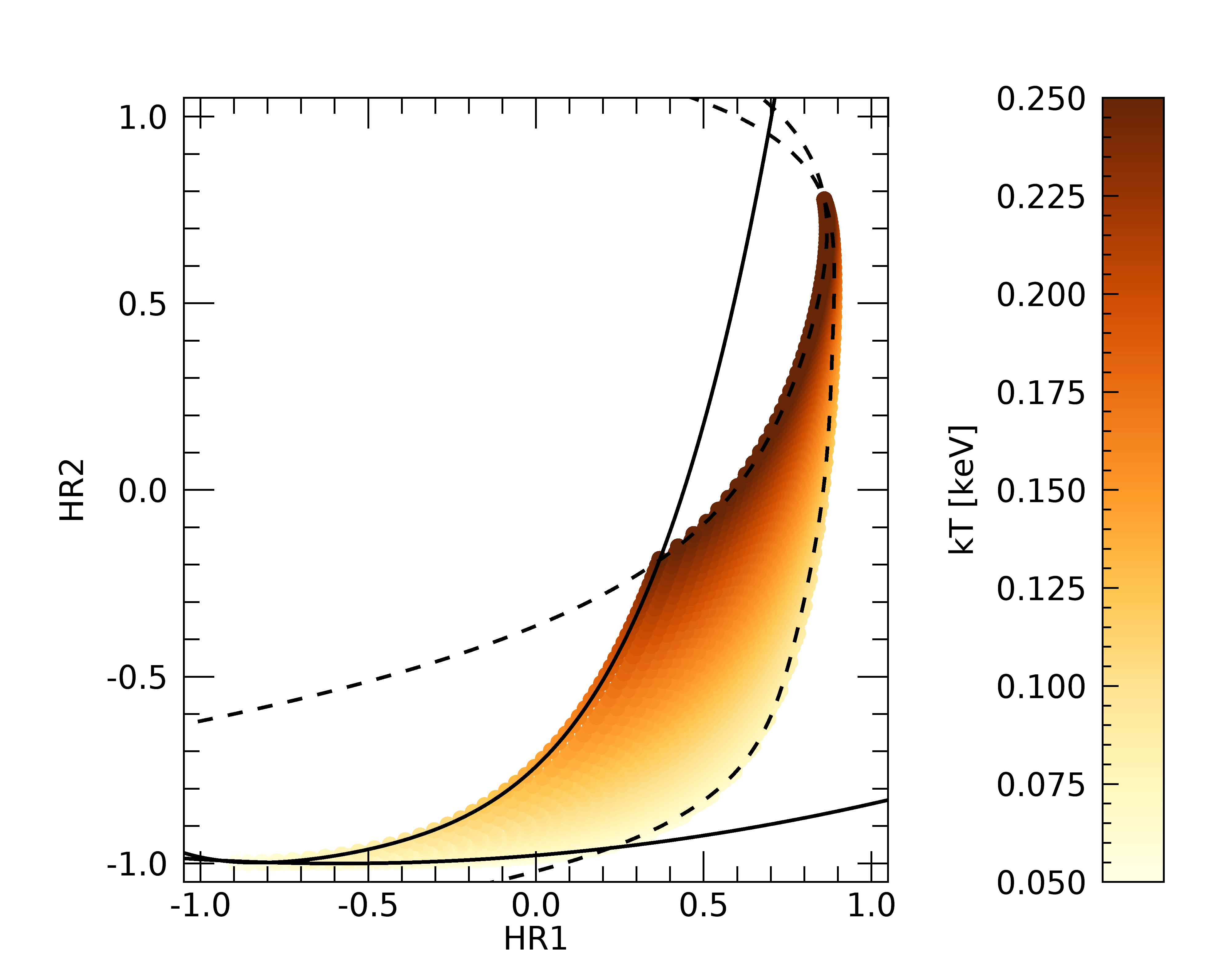}
  \includegraphics[width=0.49\textwidth]{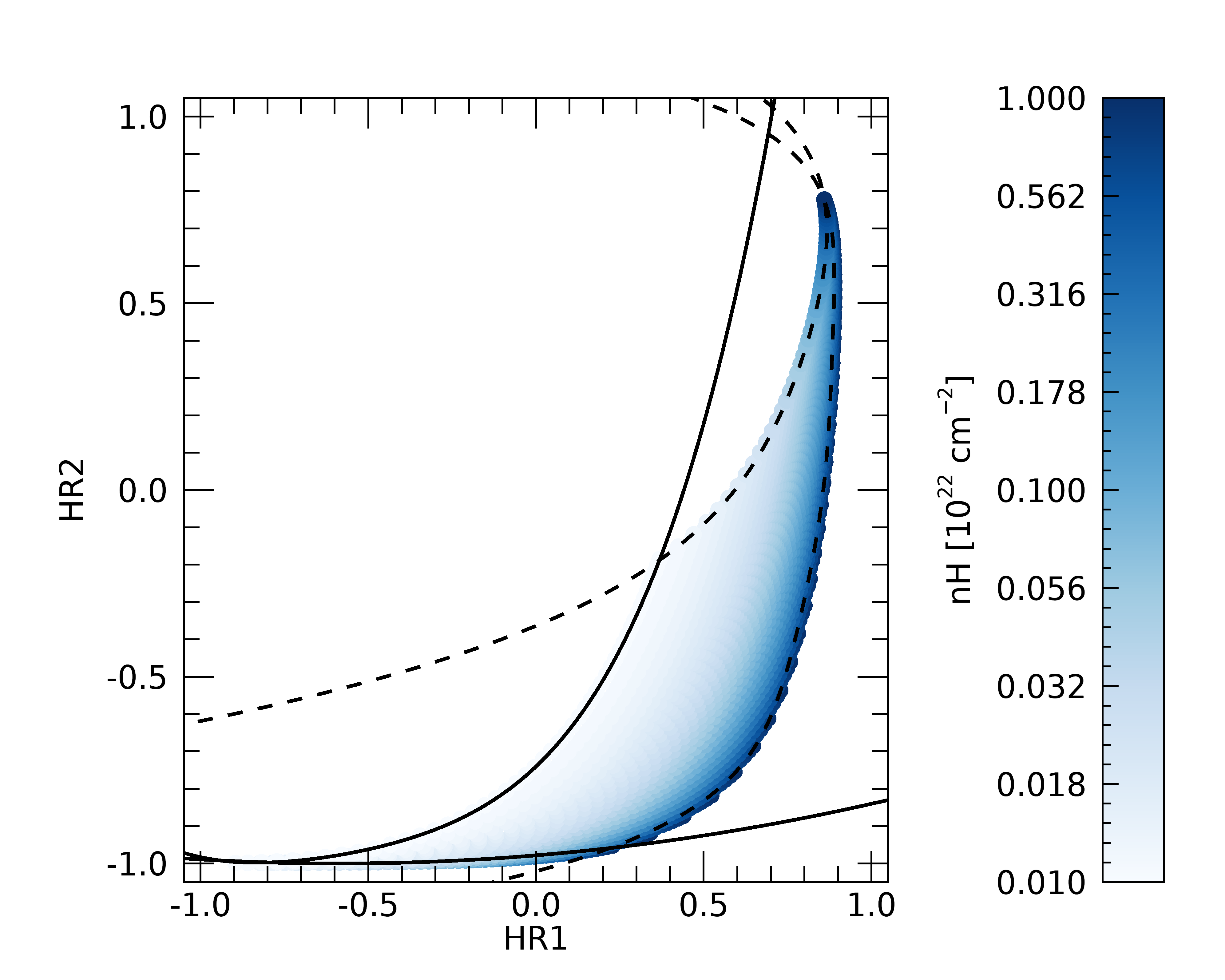}
  \caption{\hr plane of EPIC-pn sources computed with Full Frame operating mode, thin filter and on-axis. The left panel shows how HRs change varying the blackbody temperature from 0.05 to 0.25 keV; The right panel shows how HRs change varying the   absorption   from $10^{20}$ to $10^{22}$ \col.}
  \label{fig:hrktnh}
\end{figure*}

Figure~\ref{fig:hrktnh} shows the \hr plane for the thin filter.
It is clear that, for the adopted model and range of parameters, only a specific region is occupied (HR$_2<$ HR$_1$), and that to each ($kT$, \nh) pair corresponds a (HR$_1$, HR$_2$) pair. The temperature increases from low HR$_2$ to high HR$_2$ (left panel of Figure~\ref{fig:hrktnh}), while the absorption increases from low HR$_1$ to high HR$_1$ (right panel of Figure~\ref{fig:hrktnh}).

We performed  polynomial fits to the curves that include the interesting region characterizing soft sources: The curves delimiting the left and the right sides were obtained varying $kT$ and keeping \nh fixed to $10^{20}$ and to $10^{22}$ \col, respectively; those for the lower and the upper sides were obtained varying \nh and keeping $kT$ fixed to 0.05 and 0.25 keV, respectively. 
We fitted the left and the lower sides with a function $f_j(\mathrm{HR}_1)=\sum_i c_{ij}\mathrm{HR}_1^i$, while the right and the upper sides with $f_j(\mathrm{HR}_2)=\sum_i c_{ij}\mathrm{HR}_2^i$, where $j$ refers to the different side and $c_{ij}$ is the polynomial coefficient with $i$ varying from 0 to 6 at maximum. All the derived coefficients  are listed in Table~\ref{tab:poly}. 

We performed the same kind of analysis on the HR$_3$--HR$_4$ plane, and we found that the numerical HRs are restricted in the lower left corner of the plane. Considering an error of 0.05, we obtained consistent values of maximum HRs for the three filters:
$\mathrm{HR}_{3,\rm~max} = -0.78(5)$,
$\mathrm{HR}_{4,\rm~max} = -1.00(5)$.

\setlength{\tabcolsep}{0.5em}
\begin{table}
\caption{Coefficients of the polynomial fits of the four sides for each filter. }
\centering
\scriptsize
\begin{tabular}{cc|ccccccc}
    \toprule
\footnotesize{Filter} & \footnotesize{Side} & \footnotesize{$c_0$} & \footnotesize{$c_1$} & \footnotesize{$c_2$} & \footnotesize{$c_3$} & \footnotesize{$c_4$} & \footnotesize{$c_5$} & \footnotesize{$c_6$}\\
    \midrule
        Thin &left&-0.741&0.853&1.271&1.146&0.486&-&- \\
        &lower&-0.979&0.074&0.064&-&-&-&- \\
        &right&0.857&0.133&-0.222&0.024&0.399&0.100&-0.692 \\
        &upper&0.593&0.881&-1.328&1.628&-1.042&-&- \\
    \midrule
        Medium &left&-0.776&0.760&1.201&1.138&0.543&-&- \\
        &lower&-0.979&0.074&0.065&-&-&-&- \\
        &right&0.859&0.130&-0.219&0.022&0.397&0.100&-0.688 \\
        &upper&0.598&0.850&-0.859&0.272&-&-&- \\
    \midrule
        Thick &left&-0.834& 0.592&1.038&1.238&0.690&-&-  \\
        &lower&-0.979&0.080&0.076&-&-&-&- \\
        &right&0.862&0.125&-0.207&0.022&0.360&0.103&-0.651 \\
        &upper& 0.599&0.771&-0.558&-&-&-&- \\

    \bottomrule
\end{tabular}
\label{tab:poly}

\raggedright
\small

\textbf{Notes.} Left and lower sides are obtained using the fitting function $f_j(\mathrm{HR}_1)=\sum_i c_{ij}\mathrm{HR}_1^i$; right and upper sides using $f_j(\mathrm{HR}_2)=\sum_i c_{ij}\mathrm{HR}_2^i$.
\end{table}

\section{Selection of INS candidates}\label{sec:catalog}

Among all 849,991 detections contained in the \dr catalog, we first excluded spurious detections (\textsc{sc\_sum\_flag}$<$4) and those observations in which the pn was not operating. This reduced the total sample to $688,081$ detections, corresponding to $496,716$ unique sources.
Then, we excluded  $60,914$  sources that were flagged as extended  (\textsc{sc\_ext\_ml}$<$4) or as variable within the single observations  (\textsc{sc\_var\_flag}$\neq$"T").

In order to obtain reliable estimates of the source spectral shape, among the remaining sources,  we retained only the detections  with   small HR errors, namely \textsc{pn\_hr1\_err}$\leq$0.1 and \textsc{pn\_hr2\_err}$\leq$0.1.
This  significantly reduced the number of  selected detections: $34,141$, corresponding to $24,961$ unique sources.

We searched for optical and/or infrared counterparts of the selected sources in the following catalogs: USNO A2.0 \citep{USNOA}, USNO B1.0 \citep{USNOB}, GAIA DR2 \citep{GAIA}, SDSS DR12 \citep{SDSS}, Pan-STARRS1 \citep{2016arXiv161205560C} and 2MASS \citep{2MASS}. 

We also considered the work of \citet{2017ApJS..228....5K}, which catalogs point-like sources in the vicinity of nearby galaxies ($<$1.9 Mpc) observed by the \textit{Spitzer Space Telescope} (3.6--8 $\mu$m and 24 $\mu$m, \citealt{2004ApJS..154....1W}). This allowed us to identify mid-IR luminous stars in the crowded and dusty disks of large star-forming galaxies, that are not detected in the near-IR and optical bands.

The cross-correlation task was performed using the online \textit{CDS X-Match  Service}\footnote{http://cdsxmatch.u-strasbg.fr/}. The \dr catalog provides for each source the radius of the statistical error region and a systematic error, that are  added in quadrature to give the total positional uncertainty.
Based on this uncertainty and on the positional errors of the used catalogs, we adopted a threshold in the correlation radius corresponding to a significance of  $3\sigma$, based on the prescription of \citet{2011A&A...527A.126P}.

In order to estimate the X-ray to optical/IR flux ratios,  we computed the optical and IR fluxes from the magnitudes in the $R$ ($600-750$ nm) and $K_s$ ($2.0-3.0$ $\mu$m) bands, following \citet{1988ApJ...326..680M} and \citet{2003AJ....126.1090C}, respectively:
\begin{equation}
 \log{F\pdx{O}}=-\frac{m_R}{2.5}-5.37
 \label{eq:fop}
\end{equation}
\begin{equation}
 \log{F\pdx{IR}}=-\frac{m_{K_s}}{2.5}-6.95.
 \label{eq:fir}
\end{equation}
As already noticed by \citet{2012ApJ...756...27L}, AGNs and stars are empirically separated by $\log{F\pdx{X}/F\pdx{IR}}\approx -1$. 
We are interested in sources with $\log{F\pdx{X}/F\pdx{O}}$ or $\log{F\pdx{X}/F\pdx{IR}}$ larger than 3, which do not appear in our sample due to the limiting magnitudes of the adopted catalogs. 
After removing the sources with an optical and/or IR counterpart, we are left with $3,755$ detections, corresponding to $2,290$ unique sources.

As a final step of the filtering process we wanted to select the softest sources, contained in the HR region shown in Figure~\ref{fig:hrktnh}, where for each observation we used the boundaries corresponding to the appropriate filter (see Table~\ref{tab:poly}). 
We added in quadrature a systematic value of 0.05 to the statistical errors on HR  to take into account the uncertainties in the predicted HR values described above. Sources were retained if their HR$_1$ and HR$_2$  were inside the boundaries and HR$_3$ $<$ HR$_{3,\rm~max}=-0.78(5)$ considering their 1 $\sigma$ errors.

We did not use HR$_4$ because such soft sources are not detected or have a very low signal to noise ratio above 2 keV.
In this way we obtained $469$ detections, corresponding to $140$ unique sources.
In Table~\ref{tab:catlist} we summarize the steps of our filtering process of the \dr catalog.

\begin{table}
\caption{Summary of the filtering process.}
\centering
\footnotesize
\begin{tabular}{lcc}
\toprule
Filter & N. detections & N. sources\\
\midrule
Total          & 849,991   &  575,158 \\
EPIC-pn  & 688,081   &  496,716 \\
Point-like and non variable    & 580,604   &  435,802 \\
\textsc{pn\_hr1,2\_err}$\leq$0.1 &   
34,141   &   24,961 \\
Without optical/IR counterpart & 
3,755 & 2,290 \\
Soft sources according to Table~\ref{tab:poly}
& 469 & 140 \\

\midrule
Spurious or bright optical sources & 166 & 93\\
Known & 288 & 41\\
Unknown & 15 & 6 \\
    \bottomrule
\end{tabular}
\label{tab:catlist}
\end{table}

\section{Characterization of soft sources}\label{sec:ss}

\begin{figure*}
  \centering
  \includegraphics[width=0.49\textwidth]{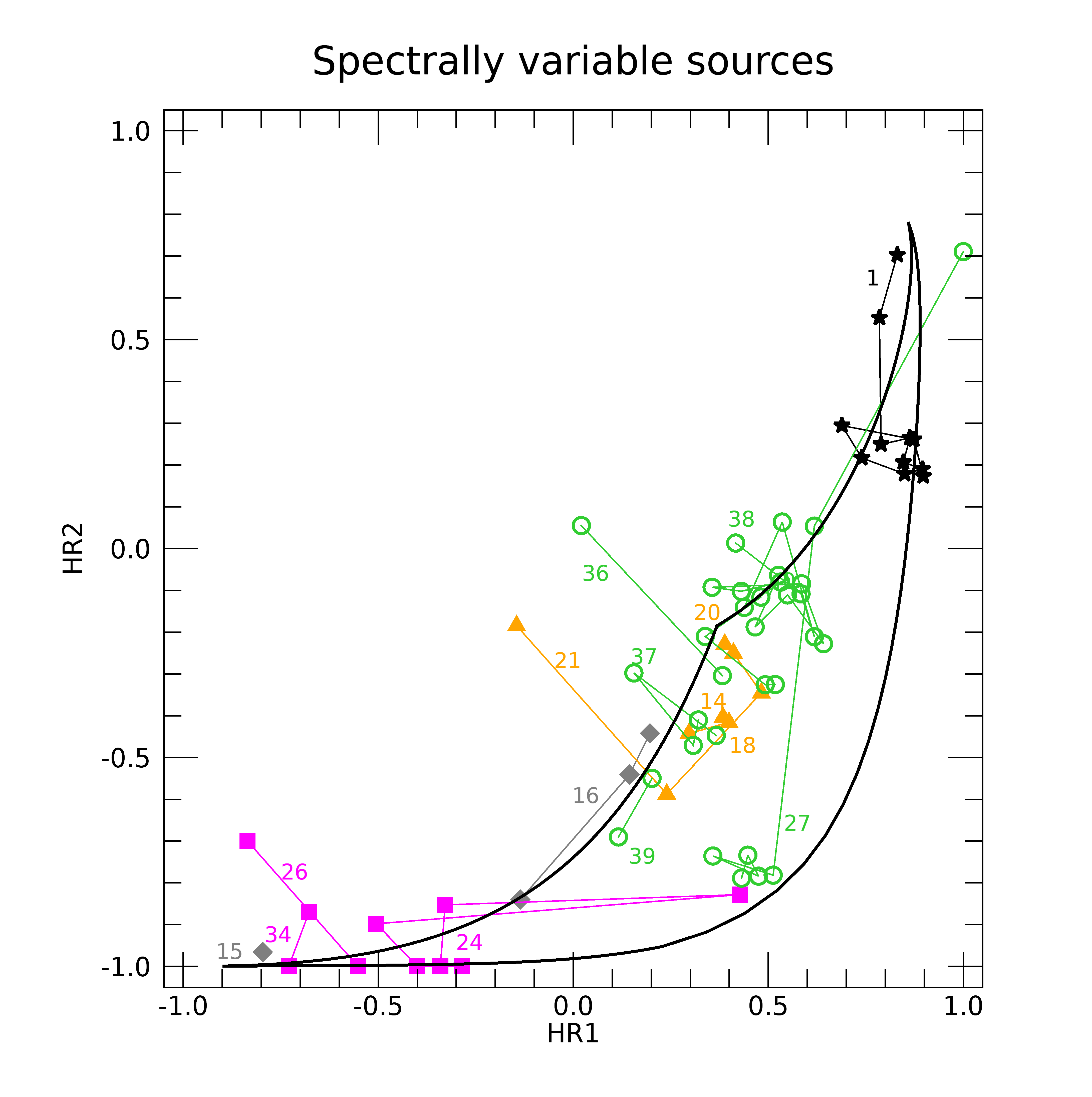}
  \includegraphics[width=0.49\textwidth]{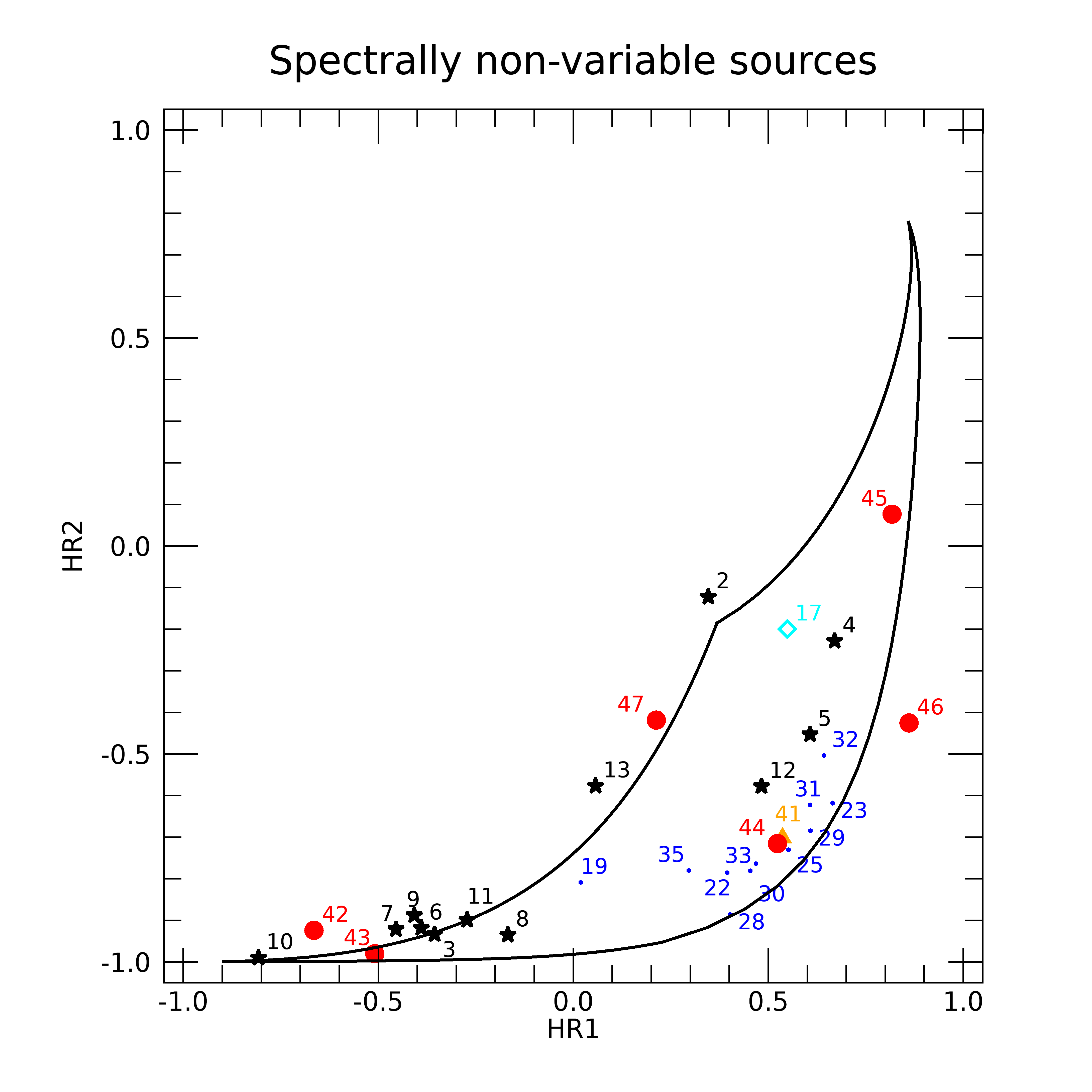}
  \caption{\hr plane of the 47 soft sources divided into spectrally variable (left panel) and non-variable (right panel). They are numbered according to Table~\ref{tab:ss}, and colored according to their class: INS (black stars), LMXB and HMXB (orange triangles), SNR (blue small dots), SSS (magenta squares), TDE (gray diamonds), AGN (cyan empty diamonds), ULX (green empty dots). The sources of unknown classification are represented by red dots.}
  \label{fig:hruk}
\end{figure*}

All the sources matching our selection criteria were checked individually in order to verify their nature. We found that a few tens of them were actually spurious detections caused by straylight from bright sources outside the field of view, or by point-like knots of extended sources such as bright SNRs. For other sources, we found plausible optical counterparts that were missed in the automatic cross-correlation due to a significant proper motion. 
We finally removed a couple of sources that lie within the soft region of the \hr plane only in one detection out of many, but had average values of HR consistently outside of the region. 

This further screening reduced the sample to 47 sources, six of which are not associated to any known X-ray object (see Table~\ref{tab:ss}). The latter, being very soft, non-variable point-like sources can be considered as potential INS candidates.

Many of the sources already identified are superimposed, or close to, nearby galaxies. To verify a possible association, we compared their position with the apparent dimensions of the galaxy,  defined by the minor and major isophotal ($m_B=25$)  diameters, and their position angle, as reported in 
The Third Reference Catalogue of Bright Galaxies \citep{1991rc3..book.....D}\footnote{See also https://heasarc.gsfc.nasa.gov/W3Browse/all/rc3.html}. Most of the sources with an associated galaxy in Table~\ref{tab:ss}, are located inside the $m_B=25$ ellipse. Only  two sources are located at much larger distances, at $\sim$3 and $\sim$5 times the isophotal radius.
The first one, 4XMM J040325.2$-$431721 is probably a background AGN \citep{2014A&A...566A.115D}, while 4XMM J031722.7$-$663704 is one of the unidentified sources (see Section~\ref{sec:j03}).

Another property that can help to classify X-ray sources is the evolution of the spectrum through different epochs. Most of the 47 selected sources were observed several times by \xmm. We divided these sources into two groups: The spectrally non-variable ones, that have consistent HR values across all observations, and the spectrally variable sources, that have significantly different HR values in different epochs. This spectral variability can be visualized  by tracks in the \hr plane.
Figure~\ref{fig:hruk} shows the HR$_1$ and HR$_2$ values of the spectrally variable (left panel) and non-variable (right panel) sources. 

We immediately notice the high correlation between the source class and its spectral variability: Among the variable sources there are extra-galactic objects as SSSs (magenta squares), ULXs (green empty dots), LMXBs (orange triangles), and one peculiar tidal disruption event (TDE, gray diamond), already noticed for its remarkably soft spectrum by  \citet{2018NatAs...2..656L}. We also included in the plot three sources (number 14, 15 and 20) even if they have only one \xmm detection, because their variability was measured by \cha and \swift-XRT. The first source, 4XMM J063045.4$-$603131, is most likely either a TDE \citep{2016A&A...592A..41M} or a galactic Nova \citep{2017AJ....153..144O}, while the other two are LMXBs in the $\omega$ Centauri globular cluster \citep{2013ApJ...763..126C} and in M31 \citep{2014ApJ...780..169B}.

Strong spectral variability is also visible in the magnetar SGR J0418$+$5729 \citep{2015MNRAS.452.3357G}. All the other INSs (black stars) have steady HRs: We found the central compact object 1E 1207.4$-$5209 \citep{2003Natur.423..725B}, the high  magnetic field pulsar J0726$-$2612 \citep{2019A&A...627A..69R}, two thermally-emitting pulsars \citep{2006ApJ...639..377M,2007ApJ...654..487N}, the INS candidate of \citet{2015A&A...583A.117P} and, as expected, six of the seven known XDINSs (the lack of RX J0420.0$-$5022 is due to the fact that its spectrum is so soft that its HR$_2$ had error $>$0.1).
We also found PSR J1400$-$1431, a binary system composed of a millisecond pulsar and a white dwarf \citep{2017ApJ...847...25S}.

The other bulk of spectrally non-variable sources is made up of SNRs (blue small dots) in nearby galaxies, as M31 and M33, the AGN candidate (cyan empty diamond) previously mentioned and one HMXB in NGC~7793 \citep{2012MNRAS.419.2095M}.
Finally, we showed with red dots the six INS candidates.

\begin{table*}
\caption{List of the soft X-ray sources}

\centering
\footnotesize
\begin{tabular}{llcccccccc}
\toprule
\midrule
Num.  & Name & Detections &  Class & Location & $\alpha_{25}~\apx{a}$ & $F\pdx{X}~\apx{b}$ & $d$ & $L\pdx{X}$  & Ref.\\
      & 4XMM &         &        &             &                      & \flux      & kpc & \lum        &     \\
\midrule
 1 & J041833.8$+$573223 & 11 &          INS/magnetar &                    - &     - &  $1.13(2)\times10^{-14}$ &       2 &    $5.4(1)\times10^{30}$ & [1]\\
 2 & J121000.9$-$522628 & 24 &          INS/CCO &                    - &     - & $2.094(2)\times10^{-12}$ &     2.1 &  $1.105(1)\times10^{33}$ & [2]\\
 3 & J072608.1$-$261238 &  1 &           INS/HB &                    - &     - &  $5.23(3)\times10^{-13}$ &       1 &   $6.26(4)\times10^{31}$ & [3]\\
 4 & J053825.1$+$281709 &  1 &          INS/RPP &                    - &     - &   $7.5(1)\times10^{-13}$ &     1.3 &   $1.52(2)\times10^{32}$ & [4]\\
 5 & J233705.7$+$615101 &  1 &          INS/RPP &                    - &     - &   $2.4(1)\times10^{-14}$ &    0.70 &   $1.40(7)\times10^{30}$ & [5]\\
 6 & J072024.9$-$312549 & 20 &        INS/XDINS &                    - &     - & $7.944(5)\times10^{-12}$ &   0.286 &  $7.775(4)\times10^{31}$ & [6]\\
 7 & J080623.3$-$412230 & 15 &        INS/XDINS &                    - &     - & $1.732(5)\times10^{-12}$ &   0.250 &  $1.295(3)\times10^{31}$ & [6]\\
 8 & J130848.1$+$212706 & 13 &        INS/XDINS &                    - &     - & $3.036(5)\times10^{-12}$ &   0.500 &   $9.08(1)\times10^{31}$ & [6]\\
 9 & J160518.4$+$324919 & 12 &        INS/XDINS &                    - &     - & $4.480(3)\times10^{-12}$ &   0.390 &  $8.152(7)\times10^{31}$ & [6]\\
10 & J185635.9$-$375436 & 39 &        INS/XDINS &                    - &     - & $5.900(2)\times10^{-12}$ &   0.123 & $1.0681(4)\times10^{31}$ & [6]\\
11 & J214303.3$+$065417 & 12 &        INS/XDINS &                    - &     - & $2.422(6)\times10^{-12}$ &   0.430 &   $5.36(1)\times10^{31}$ & [6]\\
12 & J104608.7$-$594306 &  8 &        $<$INS$>$ &                    - &     - &  $1.27(1)\times10^{-13}$ &       - &                  -  & [7][8]\\
13 & J140037.0$-$143146 &  1 &           MSP/WD &                    - &     - &  $1.01(8)\times10^{-14}$ &   0.278 &    $9.3(7)\times10^{28}$ & [9]\\
14 & J132619.8$-$472910 &  1 &             LMXB &    NGC~5139 $\apx{c}$ &     - &   $5.8(1)\times10^{-14}$ &    5.24 &   $1.90(6)\times10^{32}$ & [10]\\
15 & J063045.4$-$603113 &  1 &   $<$TDE/Nova$>$ &                    - &     - &  $2.09(2)\times10^{-12}$ &       - &                  - & [11][12]\\
16 & J215022.4$-$055109 &  3 &              TDE &        Gal1 $\apx{d}$ &     - &  $5.74(9)\times10^{-14}$ & 247,000 &   $4.19(7)\times10^{41}$ & [13]\\
17 & J040325.2$-$431721 &  1 &        $<$AGN$>$ &            NGC~1512 $\apx{e}$ & 3.019 &   $5.3(1)\times10^{-14}$ &       - &                  - & [14]\\
\midrule
18 & J004217.2$+$411537 &  2 &       $<$LMXB$>$ &                  M31 & 0.017 &  $1.11(3)\times10^{-13}$ &     778 &    $8.1(2)\times10^{36}$ & [15]\\
19 & J004239.9$+$404320 &  8 &        $<$SNR$>$ &                  M31 & 0.428 &  $5.32(7)\times10^{-14}$ &     778 &   $3.85(5)\times10^{36}$ & [15]\\
20 & J004252.5$+$411631 &  1 &             LMXB &                  M31 & 0.001 &   $3.8(1)\times10^{-13}$ &     778 &   $2.78(7)\times10^{37}$ & [15][16]\\
21 & J004315.5$+$412439 &  4 &             LMXB &                  M31 & 0.012 &   $3.1(1)\times10^{-14}$ &     778 &    $2.3(1)\times10^{36}$ & [15]\\
22 & J004339.2$+$412653 & 41 &              SNR &                  M31 & 0.029 &  $2.61(3)\times10^{-14}$ &     778 &   $1.89(2)\times10^{36}$ & [15]\\
23 & J004711.9$-$252038 &  9 &        $<$SNR$>$ &              NGC~253 & 0.187 &  $1.13(4)\times10^{-14}$ &   3,500 &   $1.66(6)\times10^{37}$ & [17]\\
24 & J005413.0$-$373308 &  6 &        $<$SSS$>$ &              NGC~300 & 1.227 &  $1.09(5)\times10^{-14}$ &   1,860 &    $4.5(2)\times10^{36}$ & [18]\\
25 & J005445.2$-$374146 &  7 &              SNR &              NGC~300 & 0.039 &  $1.76(3)\times10^{-14}$ &   1,860 &    $7.3(1)\times10^{36}$ & [19]\\
26 & J005455.0$-$374116 &  2 &              SSS &              NGC~300 & 0.001 &   $7.5(1)\times10^{-14}$ &   1,860 &   $3.12(6)\times10^{37}$ & [20]\\
27 & J022242.1$+$422402 &  7 &        $<$ULX$>$ &              NGC~891 & 0.381 &  $3.12(6)\times10^{-14}$ &  12,000 &    $5.4(1)\times10^{38}$ & [21]\\
28 & J013311.1$+$303943 & 27 &              SNR &                  M33 & 0.153 &  $1.43(4)\times10^{-14}$ &     915 &   $1.43(4)\times10^{36}$ & [22]\\
29 & J013311.7$+$303841 & 28 &              SNR &                  M33 & 0.139 &  $1.71(1)\times10^{-13}$ &     915 &   $1.72(1)\times10^{37}$ & [22]\\
30 & J013329.4$+$304911 & 15 &              SNR &                  M33 & 0.186 &  $2.29(5)\times10^{-14}$ &     915 &   $2.29(5)\times10^{36}$ & [22]\\
31 & J013331.2$+$303333 & 29 &              SNR &                  M33 & 0.047 &  $4.90(7)\times10^{-14}$ &     915 &   $4.91(7)\times10^{36}$ & [22]\\
32 & J013335.8$+$303627 & 24 &              SNR &                  M33 & 0.021 &  $1.24(4)\times10^{-14}$ &     915 &   $1.24(4)\times10^{36}$ & [22]\\
33 & J013354.8$+$303311 & 18 &              SNR &                  M33 & 0.050 &  $3.59(6)\times10^{-14}$ &     915 &   $3.60(6)\times10^{36}$ & [22]\\
34 & J013409.9$+$303220 &  2 &        $<$SSS$>$ &                  M33 & 0.121 &   $5.1(3)\times10^{-14}$ &     915 &    $5.1(3)\times10^{36}$ & [22]\\
35 & J013410.6$+$304224 & 17 &              SNR &                  M33 & 0.033 &  $1.87(4)\times10^{-14}$ &     915 &   $1.87(4)\times10^{36}$ & [22]\\
36 & J121657.0$+$374335 &  2 &              ULX &             NGC~4244 & 1.081 &  $1.02(8)\times10^{-14}$ &   4,800 &    $2.8(2)\times10^{37}$ & [23]\\
37 & J122601.4$+$333131 &  4 &              ULX &             NGC~4395 & 0.187 &  $3.79(3)\times10^{-13}$ &   4,300 &   $8.39(7)\times10^{38}$ & [21]\\
38 & J151607.2$+$561552 & 13 &        $<$ULX$>$ &             NGC~5907 & 0.476 &  $2.54(4)\times10^{-14}$ &  16,400 &    $8.2(1)\times10^{38}$ & [21]\\
39 & J223545.0$-$260451 &  2 &              ULX &             NGC~7314 & 0.633 &  $3.00(6)\times10^{-14}$ &  16,750 &   $1.01(2)\times10^{39}$ & [21]\\
40 & J231823.9$-$422354 &  4 &        $<$ULX$>$ &             NGC~7582 & 0.720 &   $7.6(3)\times10^{-15}$ &  21,200 &    $4.1(1)\times10^{38}$ & [21]\\
41 & J235800.3$-$323454 & 10 &             HMXB &             NGC~7793 & 0.285 &   $6.4(2)\times10^{-15}$ &   3,900 &   $1.16(5)\times10^{37}$ & [24]\\
\midrule
42 & J022141.5$-$735632 &  1 &                ? &                    - &     - &   $9.3(3)\times10^{-14}$ &       - &                  - & -\\
43 & J031722.7$-$663704 & 29 &                ? &            NGC~1313 $\apx{e}$ & 4.825 &   $6.7(1)\times10^{-15}$ &       - &                  - & -\\
44 & J175437.8$-$294149 &  4 &                ? &                    - &     - &   $1.6(1)\times10^{-14}$ &       - &                  - & -\\
45 & J180528.2$-$273158 &  1 &                ? &                    - &     - &   $3.9(1)\times10^{-14}$ &       - &                  - & -\\
46 & J181844.3$-$120751 &  3 &                ? &   - &     - &   $8.5(3)\times10^{-14}$ &     - &    - & -\\
47 & J220221.4$+$015330 &  1 &                ? &                    - &     - &  $1.05(2)\times10^{-12}$ &       - &                  - & -\\
\bottomrule

\end{tabular}
\label{tab:ss}

\raggedright
\small

\textbf{Notes.} Sources are listed according to their locations and their classes: Central compact object (CCO); high-B pulsar (HB); rotation-powered pulsar (RPP); X-ray-dim isolated neutron star (XDINS); millisecond pulsar (MSP); white dwarf (WD); low-mass and high-mass X-ray binary (LMXB and HMXB); tidal disruption event (TDE); active galactic nuclei (AGN); supernova remnant (SNR); super-soft source (SSS); ultraluminous X-ray source (ULX). In brackets are tentative classifications.
$\apx{a}$ Parameter computed from $D_{25}$ and $R_{25}$ to determine if the source is within the isophotal ellipse.
$\apx{b}$ Flux in 0.2--2 keV reported in the \dr catalog and calculated assuming a power-law spectrum ($\Gamma=1.42$; \nh $= 1.7\times10^{20}$ \col). In case of sources with multiple detections, we reported the average value. 
$\apx{c}$ Globular cluster also known as $\omega$ Centauri. 
$\apx{d}$ Galaxy 2MASX J21502221-0550590. 
$\apx{e}$ Due to the large value of $\alpha_{25}$ an association with the galaxy is unlikely.\\
References: [1]~\citet{2015MNRAS.452.3357G}; [2]~\citet{2003Natur.423..725B}; [3]~\citet{2019A&A...627A..69R}; [4]~\citet{2006ApJ...639..377M}; [5]~\citet{2007ApJ...654..487N}; [6]~\citet{2009ASSL..357..141T}; [7]~\citet{2009A&A...504..185Pires}; [8]~\citet{2015A&A...583A.117P}; [9]~\citet{2017ApJ...847...25S}; [10]~\citet{2013ApJ...763..126C}; [11]~\citet{2016A&A...592A..41M}; [12]~\citet{2017AJ....153..144O}; [13]~\citet{2018NatAs...2..656L}; [14]~\citet{2014A&A...566A.115D}; [15]~\citet{2011A&A...534A..55S}; [16]~\citet{2014ApJ...780..169B}; [17]~\citet{2008MNRAS.388..849B}; [18]~\citet{2014ApJ...780...39L}; [19]~\citet{2000ApJ...544..780P}; [20]~\citet{2006A&A...458..747C}; [21]~\citet{2019MNRAS.483.5554E}; [22]~\citet{2004A&A...426...11P}; [23]~\citet{2003ApJ...582..654C};  [24]~\citet{2012MNRAS.419.2095M}.
\end{table*}

\subsection{Analysis of the six unknown sources}\label{sec:unknown}

To study in more detail the six sources of  unknown nature  and located  far from  nearby galaxies that we consider as potential  INS candidates, we analyzed other observations in the \cha, \swift-XRT and \xmm  public archives.

In order to maximize the signal-to-noise ratio for these generally faint sources, we extracted their EPIC-pn spectra using the maximum likelihood method as described in \citet{2019ApJ...872...15R,2021A&A...646A.117R}. The energy bins are chosen in such a way to have a significant detection, that means at least 50 counts per bin.
Spectral fitting was done with the XSPEC software. 

For all the sources discussed below, power-law models could be rejected with high confidence either because of reduced $\chi^2 \geq 2$ or because of unphysically  large photon indices ($\Gamma \geq 6$).

\begin{table*}
  \caption{Spectral properties of the six unknown sources}
  \centering
  \footnotesize
  \begin{tabular}{lccccccccc}
\toprule
\midrule
Name & Spectral &\nhmax$\apx{a}$ & \nh & $kT$ & $R\pdx{BB,~1\,kpc}$ & $F\pdx{X}$ & $m_R$ & $\chi^2/\,$dof & nhp\\
4XMM & model & $10^{21}$ \col& $10^{21}$ \col & keV & km & \flux & $\apx{b}$ & & \\
\midrule
J022141.5$-$735632 & BB & 1.4 & $0.4\pm0.1$ & $0.062\pm0.004$ & $6.7_{-1.7}^{+2.5}$ & $(4.0_{-1.0}^{+1.5})\times10^{-13}$  & $31.7_{-0.6}^{+0.7}$ & 23.10 / 16 & 0.111\\[3pt]

J022141.5$-$735632 & PL+BB & 1.4 & $0.5\pm0.2$ & $0.056\pm0.004$ & $10.8_{-3.3}^{+5.3}$ & $(5.9_{-1.8}^{+2.9})\times10^{-13}$  & $30.8_{-0.8}^{+0.9}$ & 13.51 / 15 & 0.563\\[3pt]

J022141.5$-$735632 & BB+BB & 1.4 & $0.8\pm0.2$ & $0.049\pm0.004$ & $21.2_{-6.8}^{+12.7}$ & $(1.0_{-0.4}^{+0.7})\times10^{-12}$  & $29.5_{-0.8}^{+1.0}$ & 9.08 / 15 & 0.873\\[3pt]

J031722.7$-$663704 & BB & 0.6 & $1.2\pm0.2$ & $0.062\pm0.003$ & $3.3_{-0.7}^{+1.0}$ & $(9.6_{-2.2}^{+3.2})\times10^{-14}$  & $33.3_{-0.5}^{+0.6}$ & 112.41 / 76 & 0.004\\[3pt]

J175437.8$-$294149 $\apx{c}$ & BB & 3.9 & $3.2_{-1.1}^{+1.4}$ & $0.10\pm0.01$ & $<4.6$ & $(1.1_{-0.5}^{+1.1})\times10^{-13}$  & $>32.0$ & 8.50 / 10 & 0.580\\[3pt]

J180528.2$-$273158 & BB & 2.9 & $2.6\pm0.5$ & $0.24\pm0.02$ & $0.19_{-0.04}^{+0.05}$ & $(1.1\pm0.2)\times10^{-13}$ & $38.0\pm0.5$ & 15.62 / 17 & 0.551\\[3pt]

\midrule

J181844.3$-$120751 & APEC & 7.1 & $5.6\pm0.7$ & $0.30\pm0.03$ & - & $(1.2_{-0.4}^{+0.8})\times10^{-12}$  & - & 39.53 / 23 & 0.017\\[3pt]

J220221.4$+$015330 $\apx{d}$ & PL+BB & 0.4 & $0.5\pm0.2$ & $0.188\pm0.009$ & $0.84_{-0.09}^{+0.12}$ & $(9.1\pm0.8)\times10^{-13}$ & - & 49.41 / 40 & 0.146\\[3pt]

J220221.4$+$015330 $\apx{d}$ & BB+BB & 0.4 & $<0.3$ & $0.16\pm0.01$ & $1.2\pm0.2$ & $(9.8\pm1.1)\times10^{-13}$ & - & 43.89 / 40 & 0.310\\[3pt]

J220221.4$+$015330 $\apx{d}$ & BREMSS & 0.4 & $0.7\pm0.1$ & $0.53\pm0.03$ & - & $(2.3\pm0.2)\times10^{-12}$ & - & 49.10 / 42 & 0.210\\[3pt]

\bottomrule
  \end{tabular}
\label{tab:spec}

\raggedright
\small

\textbf{Notes.} Best-fit parameters of the EPIC-pn spectra of the listed sources. Errors at 1$\sigma$. The fluxes, corrected for the absorption, are evaluated between 0.2 and 2 keV.
$\apx{a}$ Total \textsc{Hi} column density for the source position according to the sky map of \citet{2016A&A...594A.116H}.
$\apx{b}$ Computed extrapolating at optical wavelengths ($\lambda=700$ nm) the best-fitting X-ray blackbody.
$\apx{c}$ A \cha ACIS-I spectrum is jointly fitted.
$\apx{d}$ A \cha ACIS-S and a \swift-XRT spectra are jointly fitted. Given that the source flux is variable, we reported the ones measured from the \xmm detection.

\end{table*}

\subsection{4XMM J022141.5$-$735632}\label{sec:j02}

The source 4XMM J022141.5$-$735632 was detected only once by \xmm on 2012 February 10 (obs.ID 0674110401), in an observation of the Magellanic Bridge, an \textsc{Hi} gaseous structure connecting the Small and the Large Magellanic Clouds \citep{1963AuJPh..16..570H}.

The fit with an absorbed blackbody is in reasonably good agreement with the data ($\chi^2_\nu=1.44$ for 16 dof, nhp = 0.11), yielding temperature $kT=0.062 \pm 0.004$ keV, unabsorbed 0.2--2 keV flux of $(4.0_{-1.0}^{+1.5})\times 10^{-13}$ \flux, and \nh $=(4\pm1)\times 10^{20}$ \col. This \nh value is significantly smaller than the total Galactic column density in this direction,  \nhmax $=1.4\times10^{21}$ \col \citep{2016A&A...594A.116H},  suggesting that this source is a nearby Galactic object.  
For an assumed distance of 1 kpc, the blackbody emitting radius is $R=6.7_{-1.7}^{+2.5}$ km and the luminosity is $(4.8_{-1.2}^{+1.8})\times 10^{31}$ \lum.
Other fits with single component models gave worse $\chi^2$ values.

A better fit can be obtained adding to the blackbody a second spectral component, such as a power law or a hotter blackbody, but, owing to the small number of counts  above 1 keV, we had to fix the photon index or the second temperature to reasonable values in order to constrain the other parameters.
In the case of the addition of a power law with photon index $\Gamma=3$, the blackbody has $kT=0.056 \pm 0.004$ keV, $R=10.8_{-3.3}^{+5.3}$ km, and  $F=(5.9_{-1.8}^{+2.9})\times 10^{-13}$ \flux ($\chi^2_\nu = 0.9$ for 15 dof). In the case of a hot blackbody with $kT\pdx{hot}=0.15$ keV, we obtained $kT=0.049 \pm 0.004$ keV, $R=21_{-7}^{+13}$ km, and  $F=(10_{-4}^{+7})\times 10^{-13}$ \flux ($\chi^2_\nu = 0.6$ for 15 dof). In both cases, the fitted absorption column is smaller than \nhmax and the spectral parameters are plausible for a INS at $\sim$1 kpc.

On the other hand, if we assume that the source is in the Magellanic Bridge, its luminosity would be of the order of $10^{35}$ ($d/60$ kpc)$^2$ \lum. Although this luminosity is consistent with that of an X-ray binary, the absence of an optical counterpart and lack of variability (see below) disfavor this possibility.

4XMM J022141.5$-$735632 was in the field of view of \swift-XRT seven times, between 2011 and 2015. 
The individual measurements are consistent with a constant count rate of $(2.23\pm0.55) \times 10^{-3}$ cts s$^{-1}$, which, assuming the best fit parameters of the single-blackbody fit, corresponds to a flux of $F=(2.3\pm0.6)\times 10^{-13}$. The light curve shown in Figure~\ref{fig:LC} does not give evidence for significant variability on long timescales.

\subsection{4XMM J031722.7$-$663704}\label{sec:j03}

4XMM J031722.7$-$663704 is the softest source of our sample, with almost no signal above 1 keV. It was observed many times by \xmm, \cha and \swift-XRT because it is located close to the galaxy NGC~1313, which hosts two ULXs \citep{2013ApJ...778..163B} and the ultraluminous supernova SN 1978K \citep{1993ApJ...416..167R}. The angular distance of  $\sim$9'  from the center of the galaxy is about five times larger than the isophotal radius, implying that 4XMM J031722.7$-$663704 is likely not associated to NGC~1313, as also noticed by \citet{2011ApJS..192...10L}.

Only ten\footnote{obs.ID 0405090101, 0693850501, 0693851201, 0764770101, 0782310101, 0803990101, 0803990201, 0803990301, 0803990501, 0803990601.} of the about thirty available \xmm observations that span more than 17 years, have enough counts to perform a spectral analysis. We fitted  these spectra simultaneously with an absorbed blackbody, linking the column density to a common value and obtaining $\chi^2_\nu=0.80$ for 58 dof (nhp = 0.86). The best-fitting \nh is $(1.3\pm0.2) \times 10^{21}$ \col, while temperatures and emitting radii were consistent within the errors, with average values of $kT\approx0.06$ keV and $R\approx7$ km. The fluxes, represented in the lowermost panel of Figure~\ref{fig:LC} by black stars, are consistent with a constant value of $\sim$1.2$\times10^{-13}$ \flux ($\chi^2_\nu=1.17$ for a fit with a constant).

Imposing a common $kT$ gave a similar $\chi^2_\nu=0.89$ for 67 dof (\nh $=(1.3\pm0.2) \times 10^{21}$ \col and $kT=0.060\pm0.003$), while imposing the same parameters in all the spectra gave a worse fit ($\chi^2_\nu=1.48$ for 76 dof). The spectral parameters are summarized in Table~\ref{tab:spec}.

The remaining \xmm data, for which spectral analysis was not feasible, gave detections or upper limits consistent with a constant flux (see Figure~\ref{fig:LC}). We also checked that the observations obtained with \cha ($3.6\pm2.0$ net counts on 2003/10/02, obs.ID 3551) and \swift-XRT (hundreds of observations from 2006/02/03 to 2021/04/25) do not give evidence for variability.

\subsection{4XMM J175437.8$-$294149}\label{sec:j17}

The source 4XMM J175437.8$-$294149 was detected four times by \xmm and twice by \cha, in the course of a campaign started in 2004 to characterize the X-ray sources of the Galactic Bulge in a region of low extinction, called `Stanek's window' \citep{2006ApJ...647L.135V}.
Only one \cha (obs.ID 4547) and one \xmm  (obs.ID  0402280101) observations had enough counts for a spectral analysis.

We fitted the two spectra with an absorbed blackbody, imposing a common absorption value but letting the other parameters free to vary. We found a good fit ($\chi^2_\nu=0.72$ for 8 dof) with \nh $=(2.5_{-1.2}^{+1.7})\times 10^{21}$ \col, temperatures $kT\pdx{Chandra}=0.107\pm0.015$ keV, $kT\pdx{XMM}=0.120\pm0.025$ keV and fluxes $F\pdx{Chandra}=(1.3_{-0.7}^{+2.7})\times 10^{-13}$ \flux, $F\pdx{XMM}=(1.5_{-0.9}^{+4.9})\times 10^{-13}$  \flux (unabsorbed, 0.2--2 keV).

Given that temperatures and fluxes were consistent with the same values,  we fitted the two spectra linking all the parameters. This gave  \nh $=(3.2_{-1.1}^{+1.4})\times 10^{21}$ \col, $kT=0.102\pm0.014$ keV, and $F=(1.1_{-0.5}^{+1.1})\times 10^{-13}$ \flux ($\chi^2_\nu=0.85$ for 10 dof). We used these best fit parameters, also reported in Table~\ref{tab:spec}, to convert the count rates of the other \cha (obs.ID 5303) and \xmm detections (obs.ID 0206590101, 0206590201, and 0801683001) into unabsorbed fluxes. The source flux remained constant from 2004 to 2018 (see Figure~\ref{fig:LC}).

The best fit \nh is similar to the total column density in the direction of the source, \nhmax $=3.9\times10^{21}$ \col \citep{2016A&A...594A.116H}, but its relatively large uncertainty does not allow us to estimate the source distance. For a reference distance of one kpc, the blackbody normalization corresponds to an emitting radius of $\lesssim$4.6 km and bolometric luminosity of $\lesssim$10$^{32}$ \lum. This is consistent with a cooling INS at distances up to a few kpc.

\subsection{4XMM J180528.2$-$273158}\label{sec:j1805}

The source 4XMM J180528.2$-$273158 was detected in an observation (obs.ID 0305970101) of the Galactic Center region ($l=0\deg.35$, $b=-2\deg.07$). There are no X-ray observations from \swift-XRT or \cha of this sky region.

The source spectrum can be well fitted ($\chi^2_\nu=0.92$ for 17 dof, nhp = 0.55) by an absorbed blackbody with $kT=0.24 \pm 0.02$ keV and  \nh $=(2.6\pm0.5) \times 10^{21}$ \col (the total absorption in this direction is \nhmax $=2.9 \times 10^{21}$ \col, \citealt{2016A&A...594A.116H}). The 0.2--2 keV flux, corrected for absorption,  is $(1.1 \pm 0.2)\times 10^{-13}$ \flux.
If we assume a distance in the range 1--10 kpc, the emitting radius would be $R\approx0.2-2$ km, smaller than the size of a neutron star, but consistent with emission from a hot spot on the star surface.

\subsection{4XMM J181844.3$-$120751}\label{sec:j1818}

The source 4XMM J181844.3$-$120751 was detected by \xmm in 2002 and in 2014 in three observations (obs.ID 0008820301, 0008820601, and 0740990101) aimed at characterizing two collinding-wind binaries in the Open Cluster NGC~6604, which is at a distance of 1.7 kpc \citep{2004A&A...420.1061D,2005A&A...437.1029D,2015MNRAS.451.1070D}. This region was never observed by \cha or \swift-XRT.

4XMM J181844.3$-$120751, located at 12$'$ from the center of the cluster, was also noticed by \citet{2005A&A...437.1029D} and \citet{2012ApJ...756...27L}, who suggested that it is a compact object, based on its soft spectrum and the absence of an optical counterpart.

None of the three spectra gave an acceptable fit with the blackbody model ($\chi^2_\nu=2.9$ for 19 dof, nhp $=3\times10^{-5}$). Also fits with other simple single-component models (power law, bremsstrahlung, thermal disk), as well as with the combination of a blackbody plus power-law, were rejected ($\chi^2_\nu>3$).

A good fit ($\chi^2_\nu=1.2$ for 19 dof, nhp = 0.22) was instead obtained with a model of collisionally-ionized diffuse gas with abundances fixed at solar values (\textsc{apec} in XSPEC). 
The best-fit \nh is $(5.5_{-0.6}^{+0.7})\times10^{21}$ \col, similar to \nhmax $=7.1\times10^{21}$ \col \citep{2016A&A...594A.116H} and to the values obtained for the two binaries in NGC~6604 \citep{2005A&A...437.1029D,2015MNRAS.451.1070D}.

The three temperatures are consistent within errors ($kT_1=0.29\pm0.03$ keV, $kT_2=0.33\pm0.04$ keV, and $kT_3=0.27\pm0.03$ keV), as well as the unabsorbed fluxes (see Figure~\ref{fig:LC}).
Imposing the same parameters in all the spectra gave a still acceptable fit ($\chi^2_\nu=1.25$ for 23 dof), with $kT=0.30\pm0.03$ keV and $F=(1.2_{-0.4}^{+0.8})\times 10^{-12}$ \flux.

\subsection{4XMM J220221.4$+$015330}\label{sec:j22}

4XMM J220221.4+015330, the brightest source of our sample, was detected three times in \swift-XRT (obs.ID 00036084001), \cha (obs.ID 10142) and \xmm (obs.ID 0655346833) short observations.
The three corresponding spectra have a good signal-to-noise ratio and clearly indicate spectral variability. 
They could not be fitted  by either a blackbody or a power law ($\chi^2_\nu>2$).
Acceptable fits were obtained adding a second component, but only when the cooler component has a normalization free to vary among the epochs.

The power-law plus blackbody best-fit parameters are  \nh $=(5\pm2)\times10^{20}$ \col, $\Gamma=3.0\pm0.3$, $N=(1.5\pm0.5)\times 10^{-4}$ photons cm$^{-2}$ s$^{-1}$ keV$^{-1}$ at 1 keV, and $kT=0.188\pm0.009$ keV  ($\chi^2_\nu=1.23$ for 40 dof, nhp = 0.15). The blackbody flux shows significant variations across the three observations: $F\pdx{XRT}=(5.6\pm1.5)\times 10^{-13}$ \flux, $F\pdx{Chandra}=(4.0\pm0.9)\times 10^{-13}$ \flux, and $F\pdx{XMM}=(9.1 \pm0.8)\times 10^{-13}$ \flux.

\begin{figure*}
  \centering
  \includegraphics[width=0.49\textwidth]{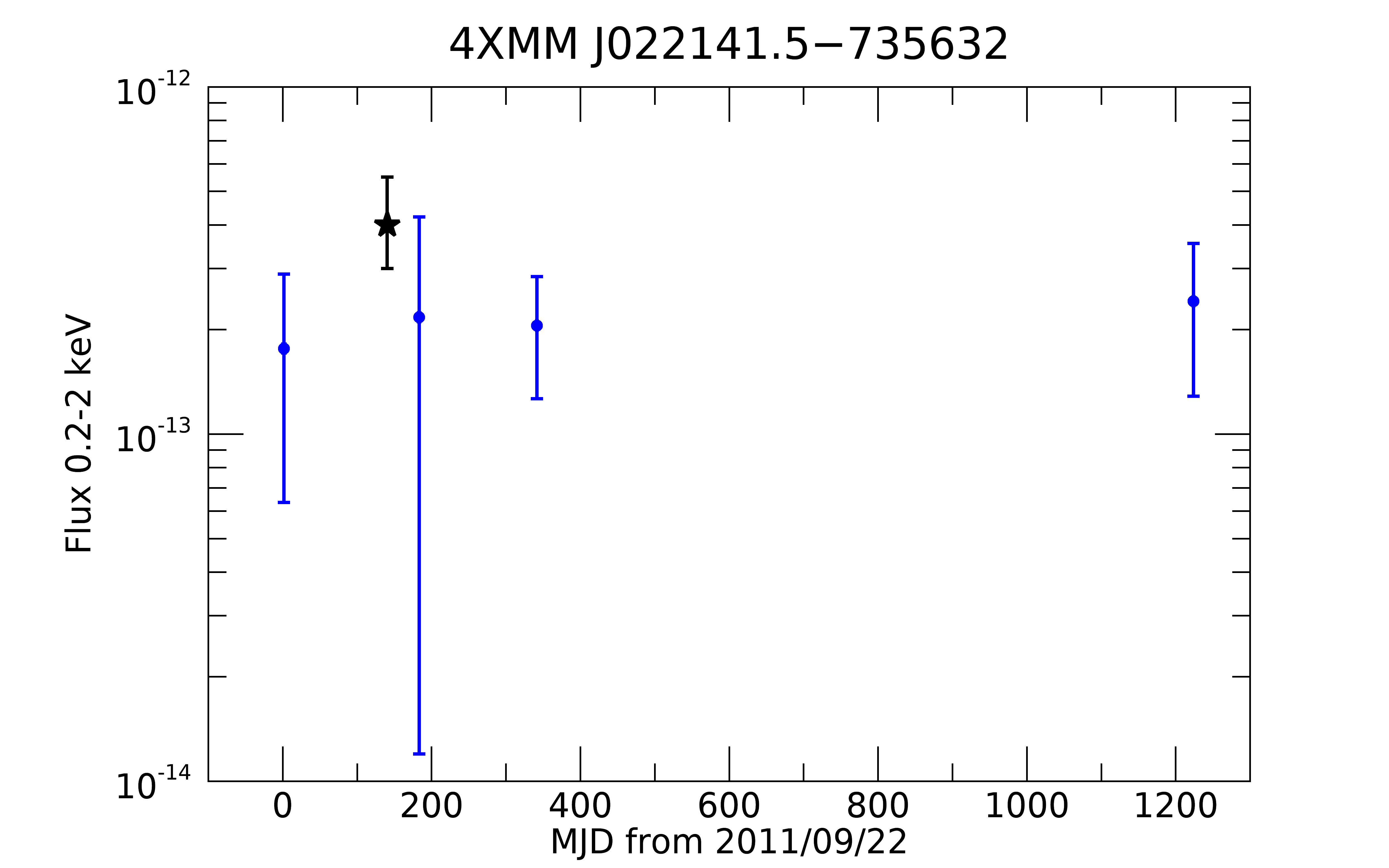}
  \includegraphics[width=0.49\textwidth]{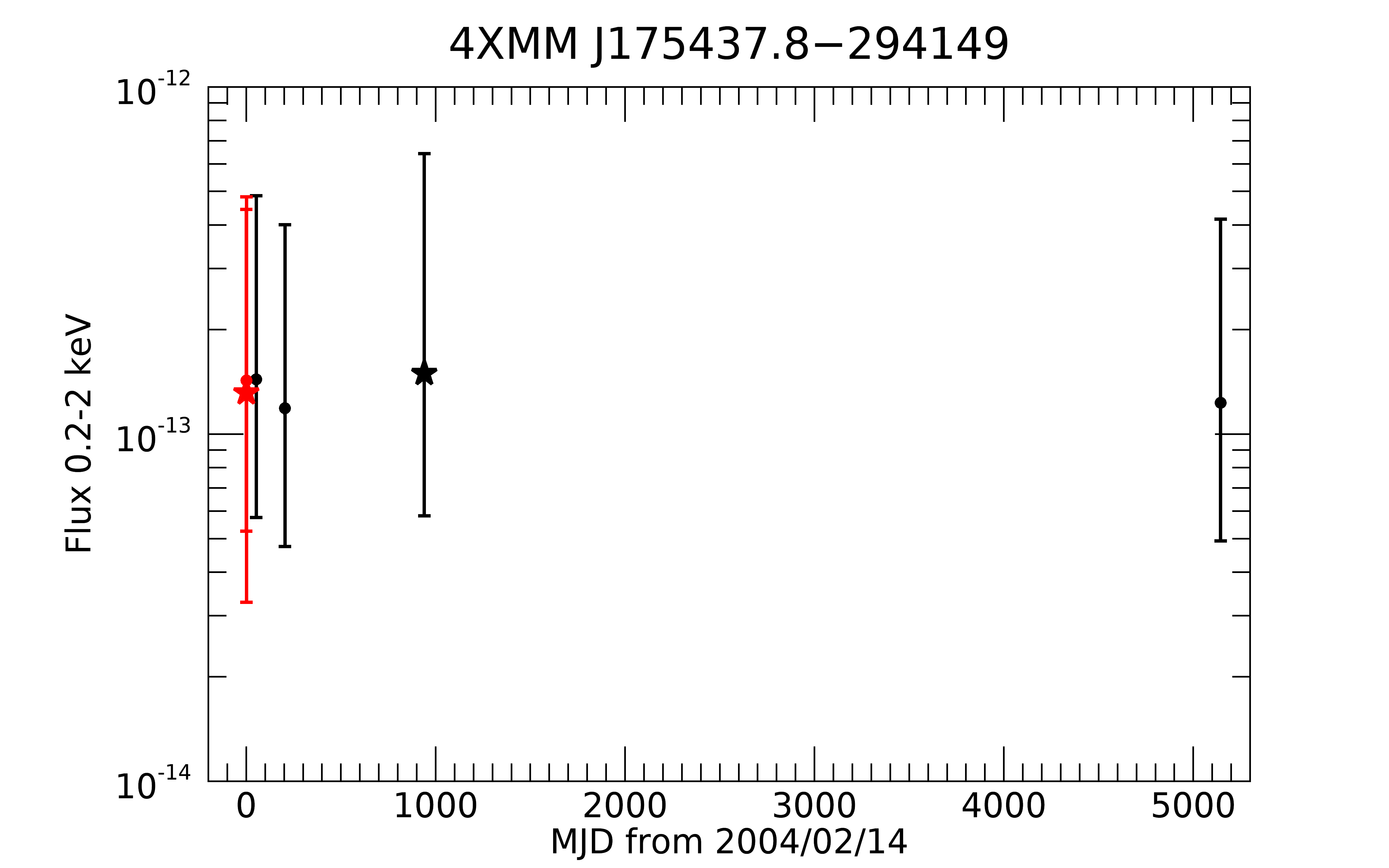}
  \includegraphics[width=0.49\textwidth]{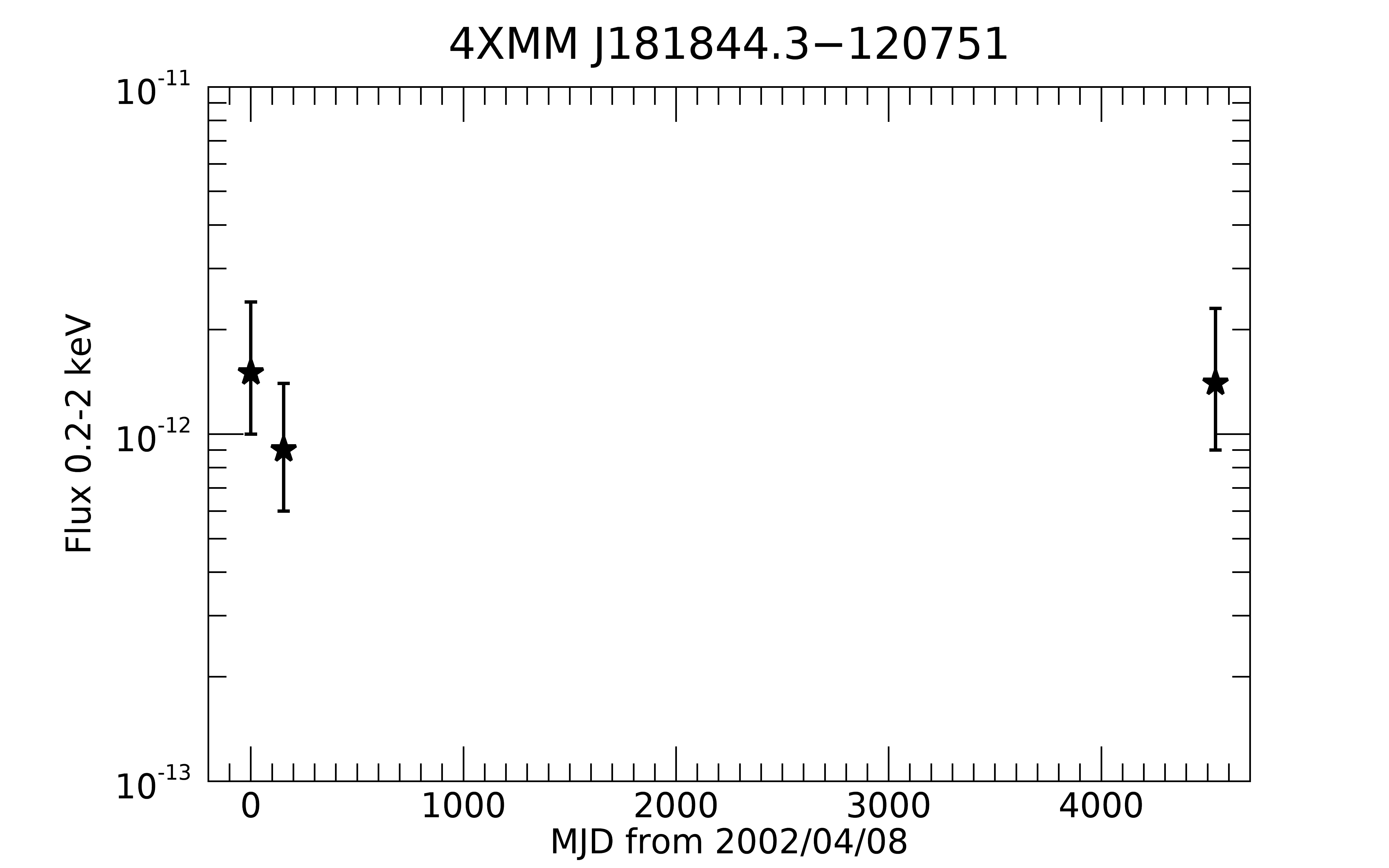}
  \includegraphics[width=0.49\textwidth]{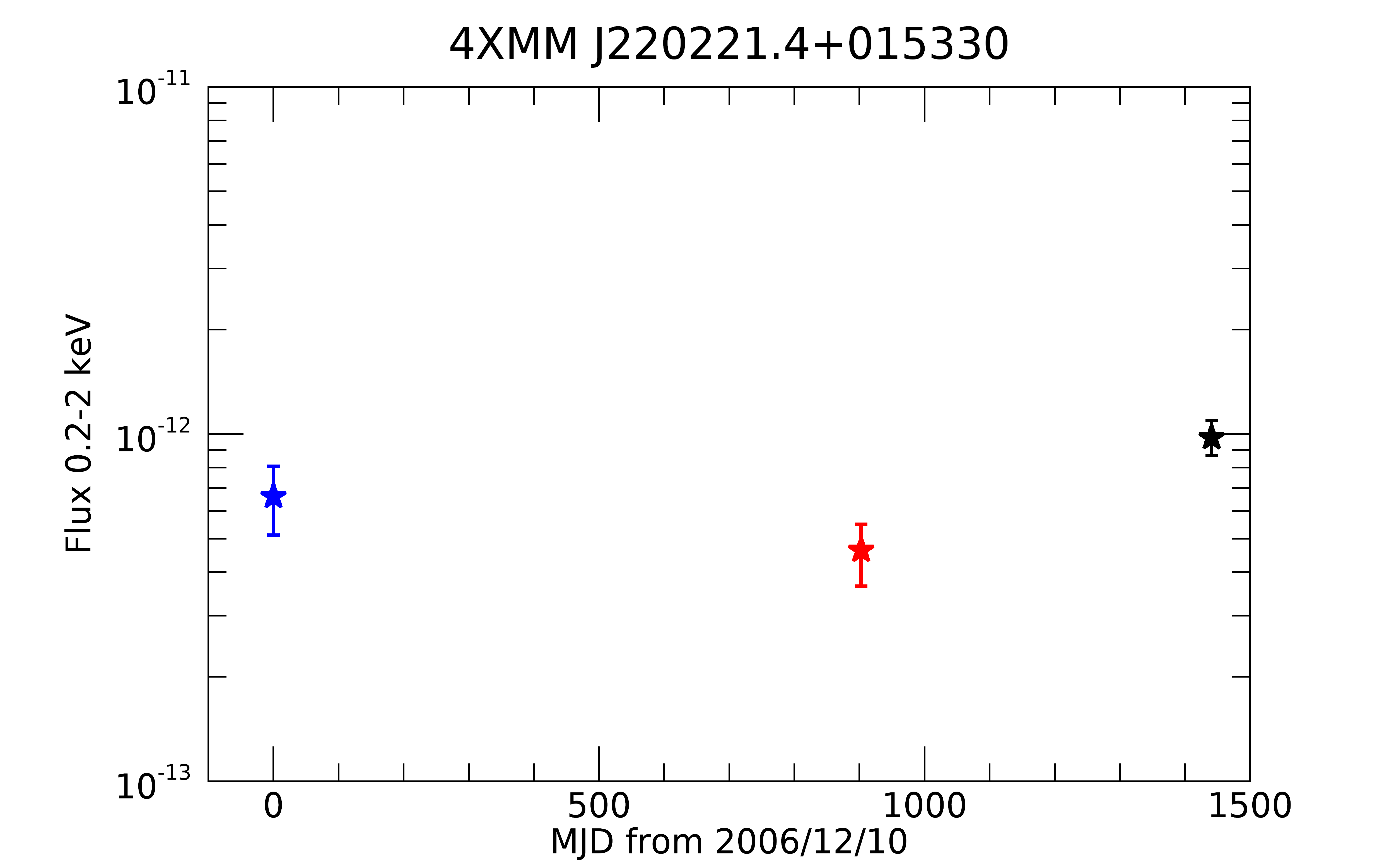}
  \includegraphics[width=1.0\textwidth]{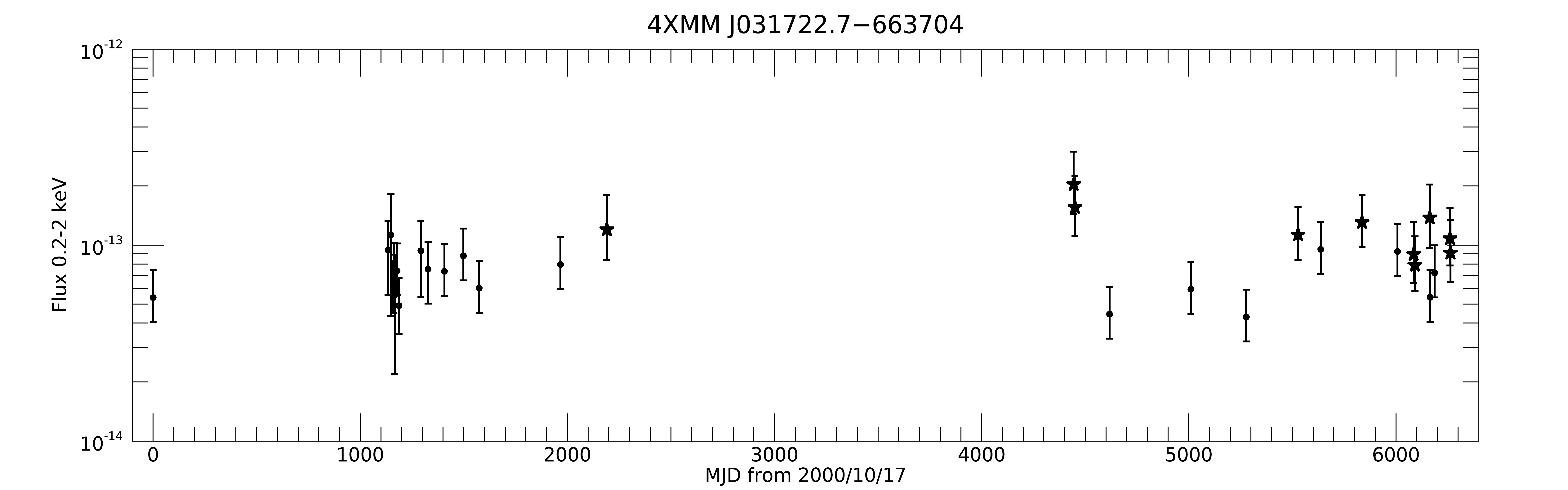}

\caption{Long-term light curves of the sources with multiple observations. The flux, corrected for absorption, is evaluated between 0.2 and 2 keV. Stars: Values measured from a spectral fit; dots: Values derived with the FTOOL \textsc{pimms} with the best fit parameters; black points: \xmm EPIC-pn; red points: \cha ACIS; blue points: \swift-XRT. 
  Top left panel: 4XMM J022141.5$-$735632; top right panel: 4XMM J175437.8$-$294149; middle left panel: 4XMM J181844.3$-$120751; middle right panel: 4XMM J220221.4$+$015330; bottom panel: 4XMM J031722.7$-$663704.}
  \label{fig:LC}
\end{figure*}

The two-blackbody model gives a better $\chi^2_\nu=1.10$ (nhp = 0.31), a small absorption column (\nh$<2.7\times10^{20}$ \col), $kT\pdx{hot}=0.33_{-0.03}^{+0.04}$ keV, and $kT=0.16\pm0.01$ keV. The hotter blackbody has a radius of  $0.16_{-0.05}^{+0.06}$ km for a  distance of 1 kpc. The longterm light curve of the colder blackbody is shown in Figure~\ref{fig:LC}. The fluxes of the cooler blackbody vary similarly to the previous case: $F\pdx{XRT}=(6.6 \pm 1.5)\times 10^{-13}$ \flux, $F\pdx{Chandra}=(4.6_{-1.0}^{+0.9})\times 10^{-13}$ \flux, and $F\pdx{XMM}=(9.8_{-1.1}^{+1.2})\times 10^{-13}$ \flux).

We notice that a good fit ($\chi^2_\nu=1.17$ for 42 dof, nhp = 0.21) can also be achieved with an absorbed bremsstrahlung, yielding \nh$=(7\pm1)\times10^{20}$ \col and $kT=0.53\pm0.03$ keV. The fluxes are higher than the previous ones, but a significant variability is still present: $F\pdx{XRT}=(1.7 \pm 0.2)\times 10^{-12}$ \flux, $F\pdx{Chandra}=(1.2 \pm 0.1)\times 10^{-12}$ \flux, and $F\pdx{XMM}=(2.3 \pm 0.2)\times 10^{-12}$ \flux.

The source  has a possible optical counterpart with magnitudes $R=25.1$ and $K_s=22.5$ detected by VIPERS-MLS \citep{2016A&A...590A.102M}. From its high X-ray flux $\sim$10$^{-12}$ \flux, we derived $\log(F\pdx{X}/F\pdx{O})= 3.4$ and $\log(F\pdx{X}/F\pdx{IR})= 4.0$, that are typical of a compact object. However, the observed variability excludes 4XMM J220221.4$+$015330 from the list of the INS candidates.

\section{Discussion and conclusions}\label{sec:disc}

We searched for new  INS candidates in  the \dr source catalog, using selection criteria based on their spectral shape, as inferred from X-ray HRs in the softest energy bands (0.2--0.5, 0.5--1, 1--2 keV), and on cross correlations with catalogs of possible optical/IR counterparts.
In particular, we identified a region in the \hr plane, assuming a blackbody spectrum absorbed by the interstellar medium folded through the response of the EPIC-pn detector, and used it to select the softest sources.

From more than half a million X-ray sources contained in the \dr catalog, we finally obtained a sample of 47 point-like sources. This sample includes about twenty SNRs or X-ray binaries located in nearby galaxies, a few AGNs and, as expected, several already-known INSs.
The remaining six sources have an unknown nature and are located far from nearby galaxies, therefore  we considered them as potential INS candidates.
A more detailed spectral and timing  analysis, using also  \cha and  \swift-XRT data, showed that two of them are unlikely to be INSs.
4XMM J181844.3$-$120751 has a spectrum  inconsistent with blackbody emission, and is instead well fit with a thermal plasma model with $kT\approx0.3$ keV. This source is close to the Open Cluster NGC~6604, but the lack of a bright optical counterpart disfavors a stellar origin.
4XMM J220221.4$+$015330 showed clear spectral and flux variability. Its faint optical counterpart implies $\log(F\pdx{X}/F\pdx{O})= 3.4$, consistent with a peculiarly soft AGN or an X-ray binary.

The remaining four sources (4XMM J022141.5$-$735632, 4XMM J031722.7$-$663704, 4XMM J175437.8$-$294149, and 4XMM J180528.2$-$273158) have soft thermal spectra and show no evidence for long-term variability. Their temperatures and emission  radii, as inferred from blackbody fits, are consistent with emission from hot spots or from (a large fraction of) the whole surface of INSs. The spectrum of the source with the highest signal to noise ratio,  4XMM J022141.5$-$735632, was better fitted with  two components  models: Either two blackbodies ($kT\pdx{hot}=0.15$ keV and $kT\approx0.049$ keV), or a power law plus  blackbody ($\Gamma=3$ and $kT\approx0.056$ keV), as typically observed in INSs. 

In Table~\ref{tab:spec} we also report the expected $R$ magnitude of these candidates, obtained extrapolating their best-fitting X-ray blackbody. These magnitudes are rather faint, but we note that these can be considered as lower limits, because the optical counterparts identified for INSs are a factor $\sim$5--10 brighter than the extrapolation of the Planckian spectra \citep{2009Msngr.138...19M}.
Whether this is due to a deviation from blackbody emission and/or a non-uniform temperature distribution, or a contamination from a nearby diffuse source, is still an open issue \citep{wan17,wan18}.
Deep multiwavelength observations are needed to identify the counterparts of these four sources, and possibly confirm the suggested neutron star nature, e.g., through the measurement of proper motions.

The small number of candidate INSs we found in the \dr catalog is not surprising, considering the rarity of thermally-emitting neutron stars sufficiently close to be detected in soft X-rays  \citep{2008A&A...482..617P}. 
The Log$N$--Log$S$ derived by these authors with a detailed evolutionary synthesis model, and taking into account the nonuniform distribution of interstellar absorption, predicts of the order of $\sim$100 cooling INSs in the whole sky above 0.001 ROSAT counts s$^{-1}$ (corresponding to a 0.2--2 keV flux of a few 10$^{-15}$ \flux). The \xmm observations used in the \dr catalog have a non uniform sensitivity, down to $F_{0.2-2}\sim10^{-16}$ \flux for the faintest detected sources. However, our requirement of having HRs with errors $<$0.1 limited our search to sources with fluxes above $\sim$10$^{-15}$ \flux. Considering that the sky coverage of the catalog is $\sim$3\% only, the small number of candidates found in our work is consistent with the expectations.
It is possible that other INSs are among the weakest \dr sources for which an accurate spectral characterization is currently impossible.

\section*{Acknowledgements}
The scientific results reported in this article are based on data obtained with \xmm, \swift, and \cha.
The data analysis has benefited from
the software provided by the NASA's High Energy Astrophysics Science Archive Research Center (HEASARC).
We extensively made use of the Simbad, VizieR and X-Match databases, operated at CDS, Strasbourg, France.
We acknowledge the whole \xmm SSC Consortium for the \dr catalog production.
We thank F. Pintore for help with the \cha data and the referee F. Walter for useful comments.
We acknowledge financial support from the Italian Ministry for University and Research through grant 2017LJ39LM ``UNIAM'' and the  INAF ``Main-streams'' funding grant (DP n.43/18).  


\section*{Data availability}

All the  data used in  this article are available in  public archives.


\bibliographystyle{mnras}
\bibliography{MN-21-2844} 

\bsp	
\label{lastpage}
\end{document}